\documentclass[titlepage,twocolumn,superscriptaddress,pre,aps]{revtex4-1}
\usepackage{amsmath}
\usepackage{amsfonts}
\usepackage{amssymb}
\usepackage{graphicx}
\usepackage{color}
\usepackage[ansinew]{inputenc}
\usepackage{helvet,times}
\usepackage{bm,textcomp}

\newcommand{\ga}{\alpha^\star}
\newcommand{\RH}{R_\mathrm{Hall}}
\newcommand{\RP}{R_\mathrm{pore}}
\newcommand{\RA}{R_\mathrm{access/2}}
\newcommand{\rp}{a}
\newcommand{\heff}{h_\mathrm{eff}}

\newcommand{\qg}{\mathcal{G}}
\newcommand{\p}{\prime}

\begin{document}

\title{Golden aspect ratio for ion transport simulation in nanopores}

\author{Subin Sahu}
\affiliation{Center for Nanoscale Science and Technology, National Institute of Standards and Technology, Gaithersburg, MD 20899}
\affiliation{Maryland NanoCenter, University of Maryland, College Park, MD 20742}

\author{Michael Zwolak}
\email[\textbf{Corresponding Author:} ]{mpz@nist.gov\\}
\affiliation{Center for Nanoscale Science and Technology, National Institute of Standards and Technology, Gaithersburg, MD 20899}

\begin{abstract}
Access resistance indicates how well current carriers from a bulk medium can converge to a pore or opening, and is an important concept in nanofluidic devices and in cell physiology. In simplified scenarios, when the bulk dimensions are infinite in all directions, it depends only on the resistivity and pore radius. These conditions are not valid in all-atom molecular dynamics (MD) simulations of transport, due to the computational cost of large simulation cells, and can even break down in micro- and nano-scale systems due to strong confinement. Here, we examine a scaling theory for the access resistance that predicts a special simulation cell aspect ratio -- the golden aspect ratio -- where finite size effects are eliminated. Using both continuum and all-atom simulations, we demonstrate that this golden aspect ratio exists and that it takes on a universal value in linear response and moderate concentrations. Outside of linear response, it gains an apparent dependence on characteristics of the transport scenario (concentration, voltages, etc.)\ for small simulation cells, but this dependence vanishes at larger length scales. These results will enable the use of all-atom molecular dynamics simulations to study contextual properties of access resistance -- its dependence on protein and molecular-scale fluctuations, the presence of charges, and other functional groups -- and yield the opportunity to quantitatively compare computed and measured resistances. 
\end{abstract}

\maketitle

\section{INTRODUCTION}
Ion transport through narrow constrictions in biological membranes permits the regulation of concentrations, known as ion homeostasis, that is vital for physiological functions of cells~\cite{hille2001,kandel2000, Doyle98-1}. Moreover, ion transport through porous inorganic membranes is of interest for technological applications -- such as DNA sequencing~\cite{Zwolak08,heerema2016} -- and industrial use~\cite{baker2000membrane,mulder2012basic}. The recent progress in the fabrication of atomically thin membranes such as graphene~\cite{garaj2010,merchant2010,Schneider2010}, MoS$_2$~\cite{heiranian2015} and hexagonal boron nitride~\cite{walker2017} opens new avenues in the field of ion transport: These membranes are excellent candidates for molecular and ionic sieves~\cite{Joshi2014,abraham2017,esfandiar2017} for desalination~\cite{tanugi2012,Surwade2015} and gas separation~\cite{jiang2009porous,kim2013selective}. Their atomic thickness provides advantages for bio-sensing, such as sequencing~\cite{clarke2009,Schneider2010,merchant2010} and protein folding~\cite{si2017nanopore}. Synthetic pores -- especially ones with atomic thickness -- also provide a testing ground for understanding biological ion channels and creating biomimetic pores. In particular, ion currents through pores of controllable size can probe dehydration~\cite{Zwolak09, Zwolak10} and directly quantify its effect on selectivity~\cite{Sahu2017NanoLett, Sahu2017Nanoscale}.

\begin{figure*}[t]
\centering
\includegraphics[width=\textwidth]{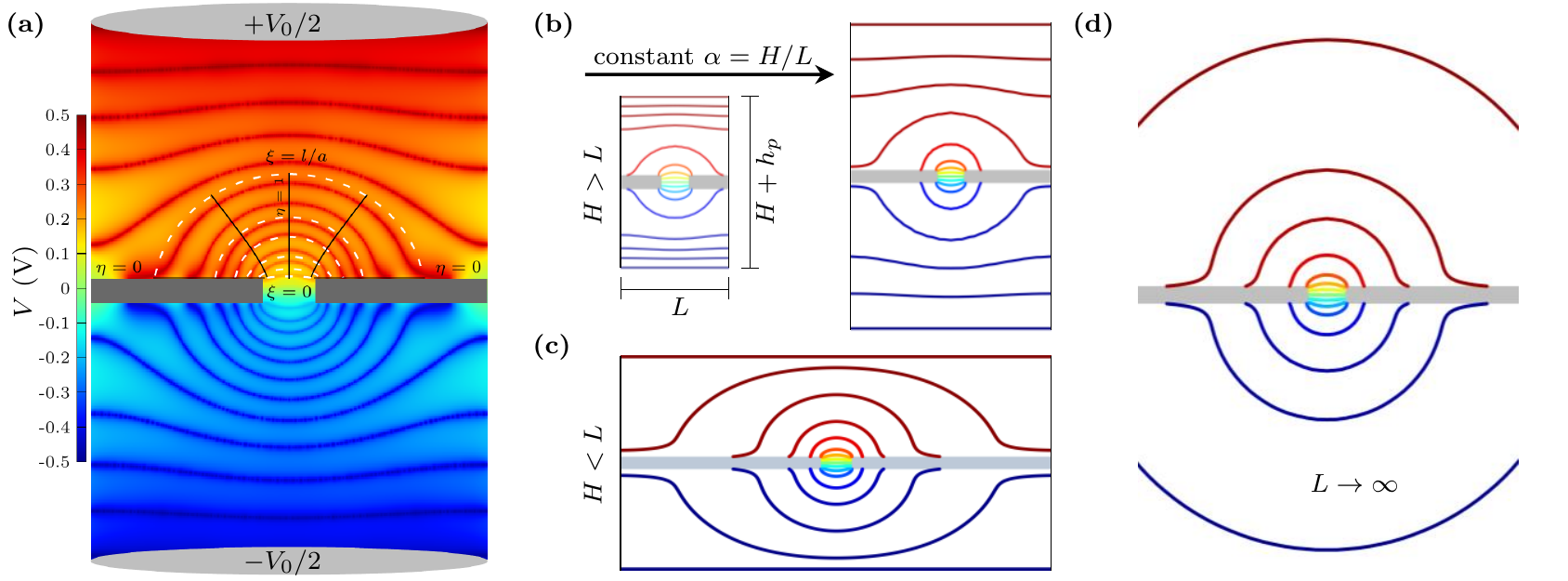} 
\caption{\label{Fig1} 
Contour plots of potential from continuum (PNP) simulations of ion transport through a nanopore. (a) Potential drop $V_0$ across the simulation cell with a pore of radius $a$ in a membrane of thickness $h_p$. The equipotential surfaces (thick, shaded red and blue lines shown on top of a heat map) are elliptical near the pore -- i.e., an access-like region. The finite size of the simulation cell curtails the access region, forcing the equipotential surfaces to transition into nearly flat profiles closer to the electrodes -- i.e., a bulk-like region. The top half shows the rotational elliptical coordinates, $\xi$ and $\eta$, we employ in modeling the access resistance. (b) Equipotential surfaces in simulation cells of increasing $L$ and $H$, with $H>L$, show the finite-size scaling of access-resistance. (c) Potential map in simulation cells with $L>H$ ($L= 32$ nm and $H=16$ nm). The equipotential surfaces are quasi-elliptical almost up to the end of the cell, but with the surfaces somewhat vertically distorted when they approach the electrode surfaces on the top and bottom. (d) As the cell size increases at a constant aspect ratio $\alpha=H/L$, more of the equipotential surfaces become ellipsoidal, i.e., access-like and the total resistance converges to its value in an infinite bulk. }
\end{figure*}

The access resistance -- part of the series resistance in patch clamp measurements -- is the resistance for ions to converge from the bulk electrolyte to the mouth of the pore~\cite{Hall1975}. It sets the upper limit of current flow through ion channels~\cite{hille1968,lauger1976} and can become the dominating resistance at low salt concentration \cite{alcaraz2017ion}. For atomically thin pores with radii sufficiently larger than the solvation shell of ions, the access resistance will be the only significant resistance, even at high salt concentration. It thus becomes an important component to understand and quantify for sensing and sequencing, which require high precision measurement and analysis of the ion current. Moreover, numerous efforts, especially using water-soluble polymers, have sought to separate pore and access contributions to the resistance in order to characterize different aspects of biological channels \cite{vodyanoy1992,bezrukov1993,bezrukov1996}.

Ideally, all-atom molecular dynamics (MD) simulations should be employed to simulate ion channels~\cite{Maffeo2012},  as only these simulations capture contextual aspects of the pores, such as molecular-scale fluctuations in pore sizes/geometries and local charges~\cite{sahu2018maxwell}. For instance, edge fluctuations and the noncircular nature of a pore in graphene, together with van der Waals interactions and dehydration, make the pore radius hard to define~\cite{sahu2018maxwell}. However, MD simulations are computationally intensive. In fact, it is often not possible to reach biologically relevant timescales using all-atom MD~\cite{eisenberg2010multiple}. Together with the long-range nature of convergence, this makes it difficult to reach the required simulation sizes to quantify the access resistance~\cite{yoo2015}. Here, we examine a scaling analysis to extract the access and pore resistances~\cite{sahu2018maxwell}, demonstrating that there is indeed a ``golden aspect ratio'' (different than the golden ratio, $(1+\sqrt{5})/2$) that removes finite size effects. We use continuum simulations in order to scan the necessary parameter space and demonstrate that the golden aspect ratio indeed removes finite size effects in MD simulation as well. 

For an infinitely large, balanced (i.e., infinite in all directions), and homogeneous system, the access resistance depends only on the resistivity of the medium, $\gamma$, and the pore radius, $a$.  Hall's expression,
\begin{equation}\label{MaxwellHall}
\RH = \frac{\gamma}{4  \rp} ,
\end{equation} 
gives the form under these idealized conditions~\cite{Hall1975}. In fact, the access resistance, Eq.~\ref{MaxwellHall}, was derived by Maxwell well over a century ago in the context of the electrical current through an orifice~\cite{maxwell_1881}. Access resistance occurs in heat flow, mass diffusion, and other related scenarios, such as electrical (e.g., disc electrodes) and thermal contacts. Hence, several authors have derived the same expression for these various physical systems, see, e.g., Gray and Mathews~\cite{GrayMathews1895}, Brown and Escombe~\cite{brown1900}, Gr{\"o}ber \cite{grober1921}, Holm~\cite{Holm1958contact}, and Newmann~\cite{newman1966}, and given various names for it (access, convergence, contact, a component of the series resistance, etc.). In the case of electronic transport, the resistance in the ballistic regime -- the Sharvin resistance \cite{Sharvin65-1} -- crosses over to Maxwell's expression when the pore is larger than the electron mean free path~\cite{wexler1966, nikolic_1999}. Although our focus is on ion transport, the general findings should be applicable to other transport scenarios as well. 

Moreover, the fact that the access resistance varies as $1/a$ rather than $1/a^2$, as it does for the pore resistance in the diffusive regime, has an interesting consequence: When the pore resistance is negligible, the current through one large pore is less than that through several smaller pores of the same total area if the pores do not interfere with each other. Nature uses this effect to maximize the gas exchange rate between the atmosphere and stomata in leaves~\cite{brown1900}, and one can envision using the same effect for maximizing permeation through atomically thin membranes, such as those through graphene and MoS$_2$, which naturally have a small pore resistance (above the dehydration limit~\cite{Sahu2017NanoLett, Sahu2017Nanoscale}). 

As we discuss later, Eq.~\ref{MaxwellHall} reflects an idealized situation. Access resistance can deviate significantly from this form when the pore and the membrane are charged~\cite{aguilella2005,luchinsky2009Self} or when only one ion species is permeable~\cite{lauger1976,peskoff1988electrodiffusion}. The effect of surface charge or concentration gradients on access resistance have been studied elsewhere~\cite{aguilella2005,peskoff1988electrodiffusion}. Here, we focus on the effect of the finite simulation cell size.
 
\section{Models and Methods} 
In addition to an infinitely large, balanced, and homogeneous bulk, the typical derivation of Eq.~\ref{MaxwellHall} assumes that a hemispherical electrode is at infinity. This means that at large distances the electric field lines extend radially outward from the pore/contact. 
%Such a directionality will emerge, though, from an infinitely large disc electrode at infinity.
Within the pore, these field lines have to transition to pointing along the symmetry axis, which we take as the $z$-axis. These symmetries can be seen from the equipotential surfaces close to the pore, as shown in Fig.~\ref{Fig1}. We will retain the ellipsoidal symmetry, which transitions from circular at the pore to hemispherical at infinity, to derive the finite-size corrections. These corrections are especially important in simulations, but can also be relevant to nanofluidic \cite{panday2016} and microelectromechanical systems (MEMS)~\cite{read2009}. 

\begin{figure*}
\includegraphics[width=0.48\textwidth]{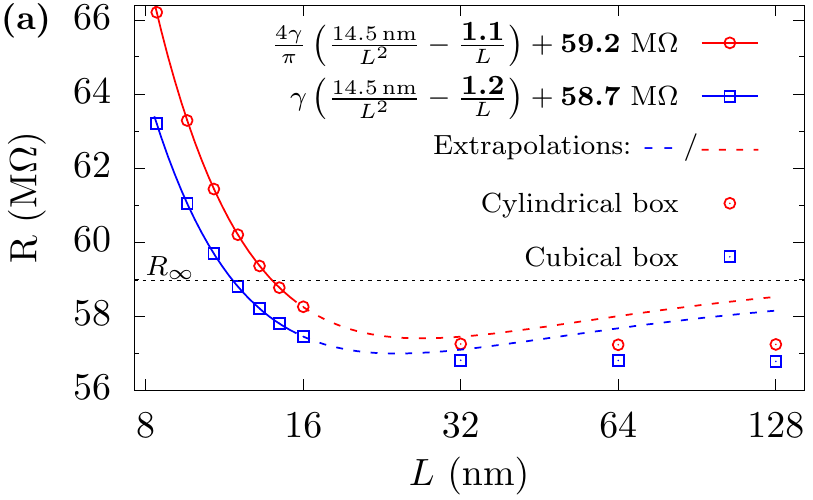} \quad
\includegraphics[width=0.48\textwidth]{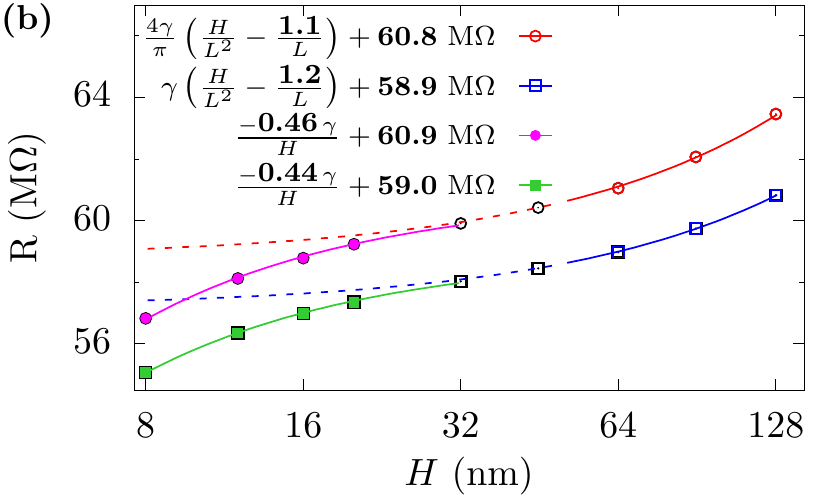} 
\caption{\label{Fig2} Variation of the resistance with changes in one of the cell dimensions. The pore radius is $a=1$ nm, the membrane thickness $h_p=1$ nm, and the resistivity $\gamma=71$ M$\Omega\cdot$nm (taken to match our all-atom MD value of 1 M KCl from prior results~\cite{sahu2018maxwell}. (a) The resistance $R$ versus cross-sectional length $L$  with electrolyte height fixed at $H=14.5$ nm. For $L \gtrsim H/f$, $R$ is smaller than the actual resistance for an infinite-size electrolyte, $R_\infty$ (horizontal dotted line). The scaling form still correctly predicts $R_\infty$ when fit for small $L$ (solid lines) but to do so it has to display non-monotonic behavior (extrapolated, dashed lines). The computed data at large $L$, however, converges to a value below $R_\infty$ as given by Eq.~\ref{HlessL}. We note that obtaining $f$ from where $R$ crosses $R_\infty$ gives about 1.0 for cylindrical and 1.2 for rectangular cells. This is in close agreement with the golden aspect ratio we obtain later, with the difference from the exact values likely due to the strong non-monotonic effects here. (b) The resistance $R$ versus total electrolyte height $H$ when cross-section is fixed at $L=50$ nm. For $H/f \gtrsim L$, $R$ increases linearly with increasing $H$ as shown by the fitted solid lines. For $H < L$, $R$ decreases with decreasing $H$ as the negative correction term $-f^\p\gamma /H$ becomes larger. The factor $f^\p \approx 0.4 $ obtained from fitting is similar to simple estimates (about 0.3). The standard error of the fits for $f$ and $R_\infty$ are less than 0.2 \%. Their fitted values are shown in bold font in the legends.}
\end{figure*} 

We have shown previously~\cite{sahu2018maxwell} how to modify Eq.~\ref{MaxwellHall} for a finite-size system [while retaining the symmetry of the problem \cite{braunovic2006,newman1966}]. This setup is easier to solve using rotational elliptic coordinates, $\xi$ and $\eta$, which relate to cylindrical coordinates, $z$ and $\rho$, via
\begin{align}
z & = \rp \, \xi \, \eta \nonumber \\
\rho & = \rp \sqrt{(1+\xi^2)(1-\eta^2)} .\label{elliptic}
\end{align}
Laplace's equation for the electric potential in this coordinate system is~\cite{abramowitz1964handbook}
\begin{equation}\label{laplace}
\frac{\partial }{\partial \xi} \left[ \left(1+\xi^2 \right) \frac{\partial V}{\partial \xi} \right]+\frac{\partial }{\partial \eta} \left[ \left(1-\eta^2 \right) \frac{\partial V}{\partial \eta} \right]=0.
\end{equation}
The boundary conditions are: (i) a constant potential at the pore mouth ($V=0$ at $\xi=0$), (ii) no radial electric field on the membrane surface ($\partial V/ \partial \eta= 0$ at $\eta=0$), and (iii) an ellipsoidal electrode of radius $l$ ($V=V_0$ at $\xi=l/\rp$). Only condition (iii) is different than that used for an infinite bulk. 

We stress that in a complex system such as ion channel, factors such as the presence of surface charges and functional groups, concentration gradients, selective ion transport, and others limit the validity of these boundary conditions. Even in a simplified system these boundary conditions only partially hold. In particular, condition (iii) is really a fictitious electrode placed at the end of the access region. We will use the location as a fitting parameter, as well as an additional contribution to capture the effect of the transition region -- the region between the access and bulk regions. Our main focus will be on obtaining the functional dependence of the resistance on cell dimensions and we will use continuum simulations to both empirically motivate and validate this dependence. 

The solution to Eq.~\ref{laplace} is, see Appendix, 
\begin{equation}
\frac{V}{V_0} =  \frac{\tan^{-1}\xi}{\tan^{-1}(l/\rp)} \label{Potential}
\end{equation}
for the potential.
%This gives the current through the pore as 
%\begin{align}
%I =& \frac{2\pi}{\gamma} \int_0^{\rp} \frac{\partial V}{\partial z}\Big|_{z=0}  \rho d\rho \,= \frac{2 \pi  \rp V_o}{\gamma \tan^{-1} (l/\rp)} .
%\end{align}
Thus, the access resistance is 
\begin{align}\label{access}
\RA &= \frac{\gamma\tan^{-1}(l/\rp)}{  2 \pi \rp}\nonumber \\ 
 %&= \frac{\gamma }{ 2 \pi \rp} \left(\frac{\pi}{2}-\frac{\rp}{l}+\mathcal{O}\left[\frac{\rp}{l}\right]^3   \right) \nonumber \\ 
 & = \frac{\gamma}{4 \rp} \left(1 - \frac{2 \rp}{\pi l }+\mathcal{O}\left[\left(\frac{\rp}{l}\right)^3\right]\right)\nonumber \\
&\approx \RH \left(1 - \frac{2 \rp}{\pi l } \right) = \RH -\frac{\gamma}{2\pi l},
\end{align} 
where we ignore higher order corrections since they are $\mathcal{O}\left[(a/l)^3 \right]$. The notation ``access/2'' indicates that this is the access resistance on a single side of the membrane. When the (simulation) cell size has, e.g., a cross-sectional length of 10 nm, $l$ has to be less than 5 nm. Hence, the expression shows that the classical form is off by $\approx$ 13 \% for a pore radius of 1 nm. This will get  worse for even moderately larger pore sizes, not to mention the difficulty in applying the Hall's form when $a$ is ill-defined.\\

\begin{figure*}[t]
\includegraphics[width=\textwidth]{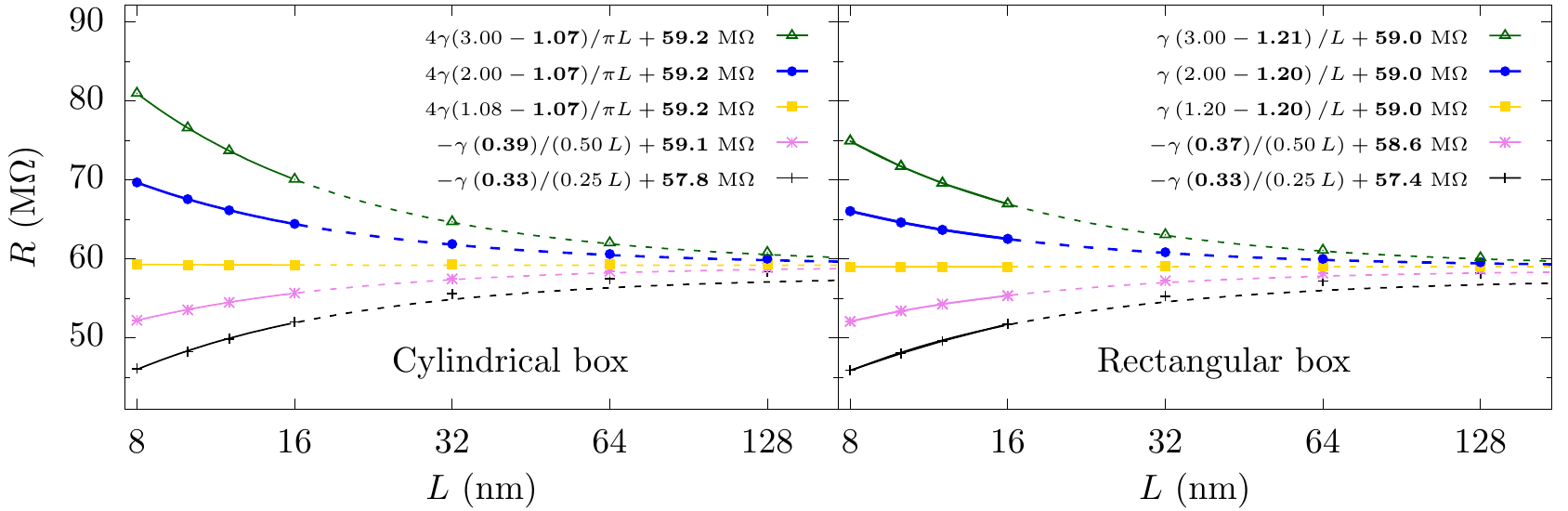} 
\caption{\label{Fig3} Resistance versus $L$ for different aspect ratios $\alpha$, with $a=1$ nm, $h_p=1$ nm, and $\gamma=71$ M$\Omega\cdot$nm. For small $\alpha$, $dR/dL>0$ and for large $\alpha$, $dR/dL<0$. At special value of $\alpha$ in between -- the ``golden aspect ratio'', $dR/dL = 0$ and the resistance is constant (i.e., no finite size effects). Here, the gold line is very close to, but not quite at, the golden aspect ratio. The solid lines show the fits $R = \frac{\gamma}{\qg L} \left( \alpha-f \right)+R_\infty$  for $\alpha>1$ and  $R = -\frac{\gamma f^\p}{\alpha L} +R_\infty$ for $\alpha < 1$,  where $\qg=\pi/4$ for cylindrical cells (left panel) and $\qg=1$ for rectangular cells (right panel). The fit parameters are shown in bold. The dashed lines extrapolate these fits, which match well with the calculations for large $L$ and yield consistent values for $R_\infty$. The performance of the scaling analysis using small $L$ indicates that the simulation cell sizes achievable with all-atom MD should be sufficient to find the total resistance (pore plus access). The error of the fits for the $f$'s and the $R_\infty$'s are about 0.5 \%  and 0.1 \%, respectively (except for $\alpha=0.25$ where the respective errors are about about 3 \% and 0.5 \% ).}
\end{figure*}

\noindent {\bf Scaling Analysis} \\
The simple boundary conditions and ellipsoidal symmetry allow us to derive the expression, Eq.~\ref{access}, for the access resistance for an idealized finite size system. However, as mentioned earlier, neither the boundary conditions nor the symmetry holds exactly in practice. In ion channels, the potential in the pore and the bulk can be coupled~\cite{luchinsky2009Self}. Nonetheless, such coupling affects the region near the pore and the boundary conditions can be taken as approximations. Another important consideration is that simulations usually have parallel disc electrodes (or homogeneous applied fields) and a uniform cross-section. Nanopore experiments and patch-clamp measurement of biological pores have even more complicated arrangements. Furthermore, in simulations, a rectangular or cylindrical cell are the natural choices. However, as shown in Fig.~\ref{Fig1}, the potential still has ellipsoidal symmetry near the pore and it starts to become flat away from the pore in the vertical direction. This is also the case in all-atom MD simulations~\cite{sahu2018maxwell}. Most importantly, therefore, we have to consider the empirical observation, see Fig.~\ref{Fig1}, that there are different electrostatic regions of the cell. In what follows, we will use this observation to develop a general scaling form. 

In the case where the bulk height $H$ (not including the membrane thickness) is greater than the cross-sectional length $L$, i.e., $H \gtrsim L$, there are three regions on each side: (a) an ellipsoidal access-like region extending from $\xi=0$ to $\xi=l/a$, (b) a flat bulk-like region extending from approximately $z=l$ to $z=H/2$ (this takes the upper membrane surface to be at $z=0$), and (c) an intermediate region between the two. When $H \gtrsim L$, the length $l$ will be some fraction of the cross-sectional length $L$ and the resistance of the intermediate region should decay as $1/L$. Thus, the total resistance is 
\begin{align}\label{scaling}
R &= R_\mathrm{access}+ R_\mathrm{bulk}+ R_\mathrm{intermediate} + \RP \\
& = 2\left(\RH - \frac{\gamma}{\pi f_1 L} \right) + \gamma \left( \frac{H- f_2L}{\qg L^2} \right)+\frac{\gamma f_3}{\qg L} +\RP  \nonumber
\end{align} 
where $\qg L^2$ is the cross-sectional area of the bulk cell ($\qg=1$ for rectangular and $\qg=\pi/4$ for cylindrical). We introduce factors $f_1$, $f_2$, and $f_3$ due to the uncertainty in the extent/contribution of these regions and the transitory nature of the boundaries between them. In Eq.~\ref{scaling}, the access region ends at $l=f_1 L/2$ and we expect $f_1$ to be $\mathcal{O}(1)$, i.e., this region encompasses a substantial fraction of the cell width (when $H \gtrsim L$). After the access region and some transition region ends in the vertical direction, the normal bulk-like region begins. It has total height $H-2 f_2 L/2$, where $f_2 L/2$ is subtracted from each side of the membrane (the membrane thickness is not included in $H$), and we expect $f_2$ to also be $\mathcal{O}(1)$.

When the cell size is infinite in all directions, the resistance reduces to
\begin{equation}
R \to R_\infty=2 \RH+\RP .\label{Rinf}
\end{equation}
Thus, from equations~\ref{scaling} and \ref{Rinf}, we obtain
\begin{align}
R 
%= & \gamma \left(\frac{H}{\qg L^2} - \frac{f_1}{\qg L}+ \frac{f_3}{\qg L} - \frac{2}{\pi f_1 L}\right)+R_\infty, \nonumber \\
 =& \frac{\gamma}{\qg} \left( \frac{H}{L^2}-\frac{f}{L} \right)+R_\infty , \label{constH}
 \end{align} 
where 
\begin{equation}
f=2\qg/\pi f_1+f_2-f_3 \label{eq:f}
\end{equation}
gives a single fitting factor. This relation shows that the finite size -- and confined -- correction to the access resistance depends on both height and the cross-section of the cell. In the context of heat flow~\cite{mikic1967thermal, cooper1969thermal}, others have shown that, for an infinitely tall cell, access resistance has functional dependence on $a/L$, which reduces to the classical form when $L \gg a$.

Equation~\ref{constH} is the resistance for $H \gtrsim L$ and its development employed both the derivation of the corrections to Hall's form and the empirical observations in Fig.~\ref{Fig1}. {\it However, Eq.~\ref{constH} does not necessarily require $R_\infty=2 \RH+\RP$, rather only that the resistance convergences to some $R_\infty$}. 
For example, $R_\infty$ can include the rectifying action of a channel~\cite{bhattacharya2011}, the effect of charges~\cite{aguilella2005,luchinsky2009Self}, or an ill-defined pore radius~\cite{sahu2018maxwell}.
Conditions such as these, ones that give an access contribution other than Hall's form, should still obey the scaling law so long as the cell is large enough to remove non-scaling finite size effects. The reason is simple, the convergence is an algebraically decaying effect, whereas structural fluctuations are completely local and charges are screened beyond few Debye lengths. This means that after some distance these features will no longer be felt by the ionic solution. We will show this explicitly in the case of non-linear response to large voltages, which is a much more drastic perturbation to the medium than fluctuations of the pore or the presence of charges.

\begin{figure*}[t]
\centering
\includegraphics{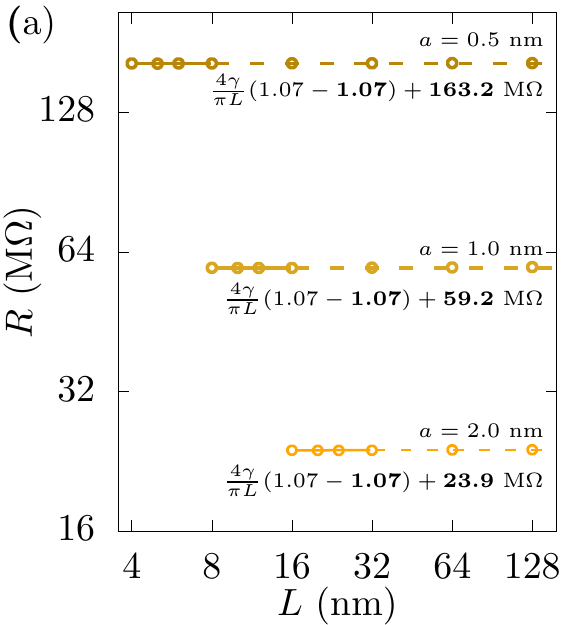}\quad
\includegraphics{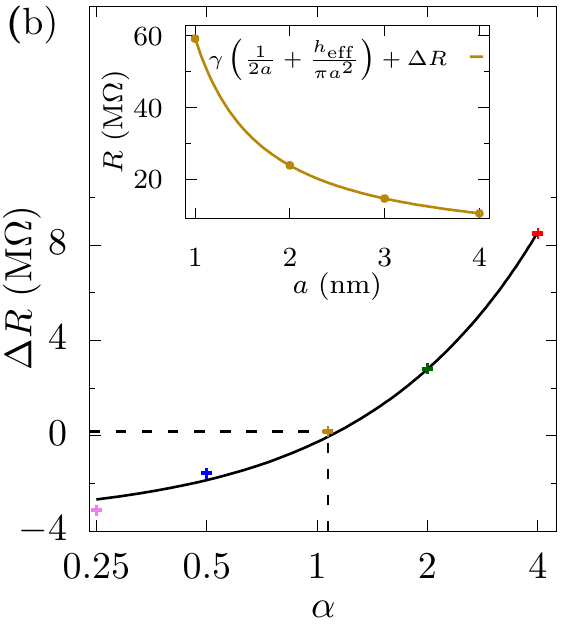}\quad
\includegraphics{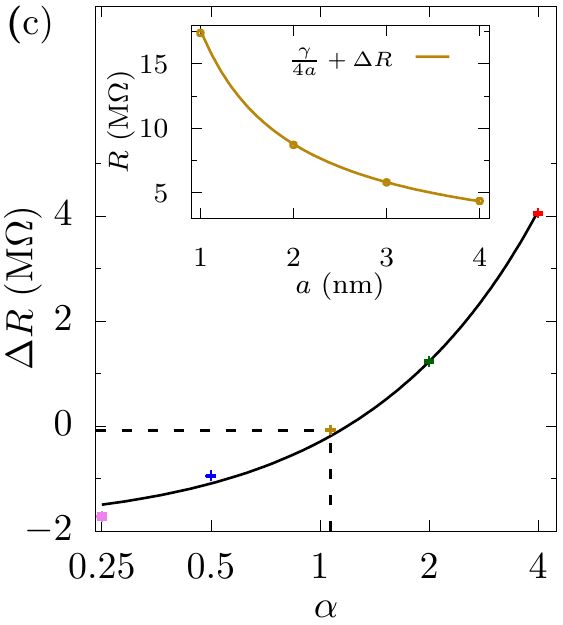} 
\caption{\label{Fig4} Universality of the golden aspect ratio with pore size. (a) The resistance versus $L$ for cylindrical cells with $\alpha=1.07$, $h_p= 1$ nm, $\gamma=71$ M$\Omega\cdot$nm, and various pore radii. The golden aspect ratio is fairly constant over this range of $a$. The solid lines show the fits $R = \frac{4\gamma}{\pi L} \left( \alpha-f \right)+R_\infty$ and the dashed lines show the extrapolations -- which make good predictions of $R$ at large $L$. The fit parameters are shown in bold. The errors of the fits for $f$'s and $R_\infty$'s are less than 0.5 \% and 0.1 \% respectively. As the pore radius increases, we examine fits for larger $L$, to roughly keep proportionality between $L$ and $a$ as there are (non-scaling) finite-size effects that will affect the simulations when the radius starts to become comparable to the cell cross-sectional length. The range of the fitting can influence the extracted $f$ (we expect that ultra-precise calculations at really large $L$ will reveal the exact golden aspect ratio, which will be close to 1.07). (b) $\Delta R$ from fits to $R = \gamma \left( 1/2a +\heff/\pi a^2\right)+\Delta R,$ for cylindrical cells of various $\alpha$, with $\heff$ and $\Delta R$ as fitting parameters. We get $\Delta R\approx 0$ for the ``golden aspect ratio" $\ga=1.07$ (dashed lines), the fit of which is shown in the inset. The effective height, $\heff$, is (5$\pm$1) \% larger than the 1 nm membrane thickness, see Fig.~\ref{Fig5}. (c) $\Delta R$ in the fits $R = \gamma/4a+\Delta R$ for half cylinders of different aspect ratios with the pore mouth set at potential $V=0$. We again get $\Delta R\approx 0$ for $\ga= 1.07$ (dashed lines), whose fit is shown in the inset.  The interpolated lines in the insets are for visual clarity only.}
\end{figure*}

For $H \lesssim L$, we will get a different form than Eq.~\ref{constH} but the considerations will be similar. Fig.~\ref{Fig1} (c) shows the equipotential surfaces for this case. The access region now ends at $l = f_1^\p H/2$, as it is not the horizontal boundary that terminates the access region but the vertical. This occurs at some fraction of the box height, $f_1^\p = \mathcal{O}(1)$, where the prime indicates it is different than the fraction in the $H \gtrsim L$ case. The normal bulk region is now almost negligible. Moreover, the cross-sectional area that contributes to the resistance in the normal bulk region is not $L \times L$, but rather should be nearly equal to $H \times H$, as only the region above the access region feeds ions into that region and contributes to the current. This is also reflected by the fact that the electric field is nearly zero for a radial distance about $H$ away from the $z$-axis. Hence, at that distance away, the electrolyte does not contribute to the resistance. We will therefore take $\gamma f_2^\p / \qg H$ for the normal bulk contribution (i.e., a height divided by a relevant cross sectional length that is proportional to $H/H^2$) and $f_2^\p$ should be very small. The transition region will also depend on $1/H$, giving
\begin{align}\label{HlessLscaling}
R &= R_\mathrm{access}+ R_\mathrm{bulk}+ R_\mathrm{intermediate} + \RP \\
& = 2\left(\RH - \frac{\gamma}{\pi f_1^\p H} \right) + \frac{\gamma f_2^\p}{\qg H} +\frac{\gamma f_3^\p}{\qg H} +\RP . \nonumber
\end{align} 
Rewriting this equation gives 
\begin{align}
R 
 =& -\frac{\gamma f^\p}{ H} +R_\infty , \label{HlessL}
 \end{align} 
 with $f^\p = 1/\pi f_1^\p - f_2^\p/\qg - f_3^\p/\qg  $. This equation entails that if we keep $H$ constant the total resistance will stay constant as we increase $L$. We estimate $f_1^\p \approx 1$ and $f_2^\p \approx f_3^\p \approx  0$ giving $f^\p \approx 1/\pi \approx 0.32$. Moreover, the correction to access resistance is negative  as it  removes part of the access region to give a lower lower total resistance.

The difficulty with the applying Eq.~\ref{constH} and \ref{HlessL} is that $R$ both depends on two geometric variables, $H$ and $L$, and it transitions from Eq.~\ref{constH} to Eq.~\ref{HlessL} at an unknown boundary (albeit $H \approx L$). On the one hand, if we hold $L$ constant and increase $H$, then $R$ will increase linearly with $H$ (e.g., this would apply both in simulations and experiments where a very narrow nanofluidic constriction leads up to the membrane/pore). This is essentially increasing the larger normal bulk region in the cell. On the other hand, if we keep $H$ constant and increase $L$, the resistance will initially decrease due to the larger cross-sectional area available for transport in the normal bulk region. However, the finite-size correction in Eq.~\ref{constH} becomes negative once $L \gtrsim H/f$. At this point, $R$ becomes smaller than $R_\infty$ and, as Eq.~\ref{HlessL} indicates, should flatten out as $L$ becomes larger. This is seen in Fig.~\ref{Fig2}.  In other words, a further increase in $L$ does not converge $R$ to $R_\infty$, because the additional cross-sectional area does not contribute to transport. One can still apply the scaling form, Eq.~\ref{constH}, so long as the fit is for a range of $L \lesssim H/f$, which Fig.~\ref{Fig2}(a) shows it extracts the correct $R_\infty$ (found by using the approach we describe below). To use the scaling form in Eq.~\ref{HlessL} one has to set $L$ to a large value and increase $H$, fitting for values $H \lesssim L$, as seen in Fig.~\ref{Fig2}(b) (one can not take $H$ near in magnitude to $L$, however, as there are corrections missing in Eq.~\ref{HlessLscaling} as $H$ approaches $L$ from below). 

In order for $R$ to converge to $R_\infty$ monotonically (in the absence of nonlinearities), whether from above or from below, both $L$ and $H$ need to increase simultaneously. An intuitive method to do so is to keep the aspect ratio, $\alpha=H/L$, constant, which simplifies Eq.~\ref{constH} and Eq.~\ref{HlessL} to
\begin{align}
R = 
\begin{cases}
 \frac{\gamma}{\qg L} \left( \alpha-f \right)+R_\infty  & \text{for } \alpha \gtrsim 1, \\
 -\frac{\gamma f^\p}{\alpha L} +R_\infty  &\text{for } \alpha < 1.
\end{cases}  \label{golden}
\end{align}
Not only do these equations give a unified scaling form $R_\infty + \mathcal{O}(1/L)$, i.e., both decay with $L$ (one from above, one from below), the former also indicates that if we choose a special aspect ratio $\ga=f$ then the finite size correction is eliminated and $R$ is independent of $L$. In other words, at this ``golden aspect ratio", $\ga$, $R=R_\infty$ for any $L$. 

It is likely not possible to find a general expression $f= 2\qg/\pi f_1+f_2 -f_3$, since the transition from access-like to bulk-like is complicated and the factors $f_1$, $f_2$ and $f_3$ will in general depend on geometric details of the pore and the boundary conditions. Nevertheless, with some reasonable approximations we can give an estimate of $f$. We expect that $f_1\approx f_2 \approx 1$ (i.e., a boundary between access and bulk-like regions at a radial distance equal to the cell edge, $l=L/2$). To estimate $f_3$, which depends on $f_1$ and $f_2$, we assume that the intermediate region has equal contributions from access- and bulk-like behavior.  This intermediate region can be assumed to have a hemispherical boundary of radius $f_1 L/2 $ and a flat boundary of cross-section $\qg L^2$ at height $f_2 L/2$ on each side of the pore. Thus the resistance of the intermediate region is estimated to be
\begin{equation}
%R_\mathrm{intermidiate} &= \frac{1}{2} R_\mathrm{access-like} + \frac{1}{2} R_\mathrm{bulk-like}  \nonumber \\
\frac{\gamma f_3}{\qg L} = \frac{1}{2} \left(\frac{2\gamma}{ \pi f_1 L}-\frac{2\gamma}{ \pi L}\right)+\frac{1}{2}\left(\frac{2\gamma f_2 L/2}{\qg L^2}\right) ,
\end{equation}
giving
\begin{equation}
f_3 = \frac{\qg}{ f_1\pi}-\frac{\qg}{ \pi} +\frac{f_2}{2}.
\end{equation}
The expression for $f_3$ likely overestimates the bulk-like contribution, while either under- or over-estimating the access contribution. 
Using it in Eq.~\ref{eq:f}, simplifies $f$ to 
\begin{align}
f &= \frac{\qg}{\pi f_1}+\frac{f_2}{2}+\frac{\qg}{ \pi}. 
\end{align}

\begin{table}[h!]
\centering
\begin{tabular*} {.8\columnwidth}{@{\extracolsep{\fill}}|l|llll|}
\hline
$f_1$ &    1.00 &    0.80 &    0.60 &    0.40 \\ \hline
$f_2$ &    1.00 &    0.80 &    0.60 &    0.40 \\  \hline
$\ga$ (rectangular) &    1.14 &    1.12 &    1.15 &    1.31\\ \hline
$\ga$ (cylindrical)    &   1.00    & 0.96  &    0.97 &    1.08  \\  \hline
\end{tabular*}
\caption{\label{Table1} Estimates of the golden aspect ratio, $\ga = f$, for various values of $f_1$ and $f_2$. We expect $f_1$ and $f_2$ to be $\mathcal{O}(1)$.}
\end{table}

Table~\ref{Table1} shows the value of $f$ for various values of $f_2=f_1$. The value of $f$ is quite insensitive to $f_1$ and $f_2$ in a reasonable range (due to the fact that one appears in a denominator and the other in a numerator), including when $f_1$ and $f_2$ are varied separately. The equipotential surfaces in Fig.~\ref{Fig1} suggest $
f_1 \approx 2/3$ and $f_2 \approx 3/4$, and thus we estimate
\begin{equation}
f = 
\begin{cases}
1.2 & \text{for rectangular box} \\
1.0 &\text{for cylindrical box}
\end{cases}  \label{GASest}
\end{equation}
for the golden aspect ratio. 

We will also examine the sensitivity of $f$ to different conditions, including whether one uses the Poisson-Nernst-Planck (PNP) equations or just Laplace's equation (i.e., Ohm's law). The latter -- the homogeneous medium approximation -- is valid when the applied potential is small and the ion concentration is large \cite{garaj2010}. Poisson's equation and the stationary Nernst-Planck equation express the spatial dependence of the potential $V$ and current density $\bf{J}$ as
\begin{equation}
\nabla^2 V = \sum_\nu \frac{q_\nu c_\nu}{\epsilon }
\end{equation}
and
\begin{equation}
{\bf J_\nu} = -q_\nu \left(D_\nu \nabla c_\nu +  \mu_\nu c_\nu \nabla V \right),
\end{equation}
where $q_\nu$, $c_\nu$, $D_\nu$, $\mu_\nu$ are the charge, concentration, diffusivity, and mobility of ion species $\nu$.  We use a commercial finite element solver for both cases. 

Finally, we expect that the ``golden aspect ratio" is applicable to MD because, at large distance and weak enough variation in fields, MD can be coarse-grain into a continuum description. We test this applicability by simulating the ionic current through graphene nanopore of radius $a=1.81$ nm immersed in 1 mol/L of KCl. We perform the all-atom MD simulation using NAMD2~\cite{phillips2005}, the CHARMM27~\cite{Feller2000} force field, and a rigid TIP3P~\cite{jorgensen1983} water model. The details of MD simulation are the same as in our previous work~\cite{sahu2018maxwell}. 

\begin{figure}
\center
\includegraphics{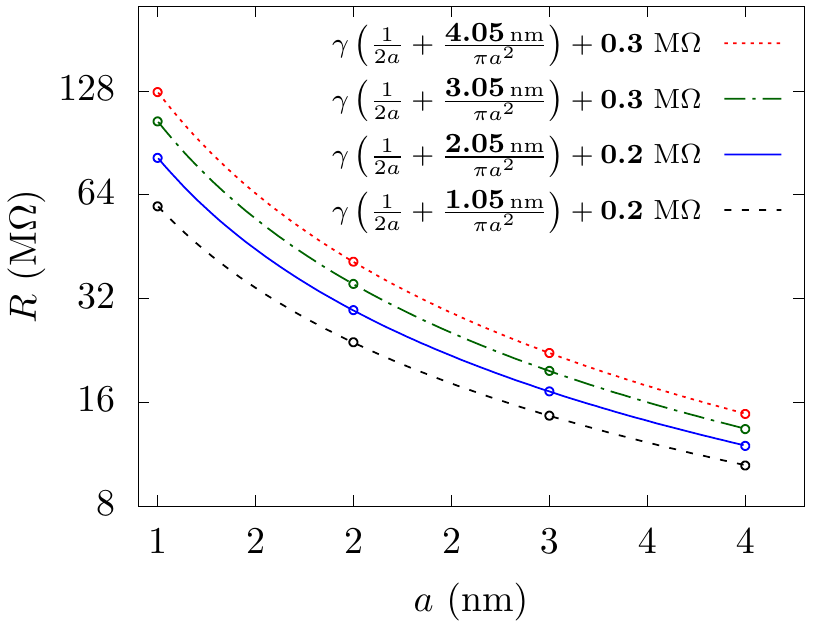} 
\caption{\label{Fig5} The resistance versus $a$ for cylindrical cells with $L=32$ nm, the golden aspect ratio $\ga=1.07$, and $h_p=$ 1.0 nm, 2.0 nm, 3.0 nm and 4.0 nm. The resistance fits into the model,  $R = \gamma \left( 1/2a +\heff/\pi a^2\right)+\Delta R $ with $\heff$ and $\Delta R$ as fitting parameters (shown in bold in the legend). The fitted $\heff$ is slightly larger than the actual membrane thickness but its relative difference diminishes as the membrane thickness is increased. This is due to the curvature of the potential at the pore mouth and not due to numerical grid-size errors (a ten times smaller grid increase this value to 7 \% and thus we expect this height correction to converge to something of this order.). The standard errors of $\heff$ and $\Delta R$ are about 0.01 nm and 0.04 M$\Omega$ in each case.}
\end{figure}

\begin{figure*}
\centering
\includegraphics[width=\textwidth]{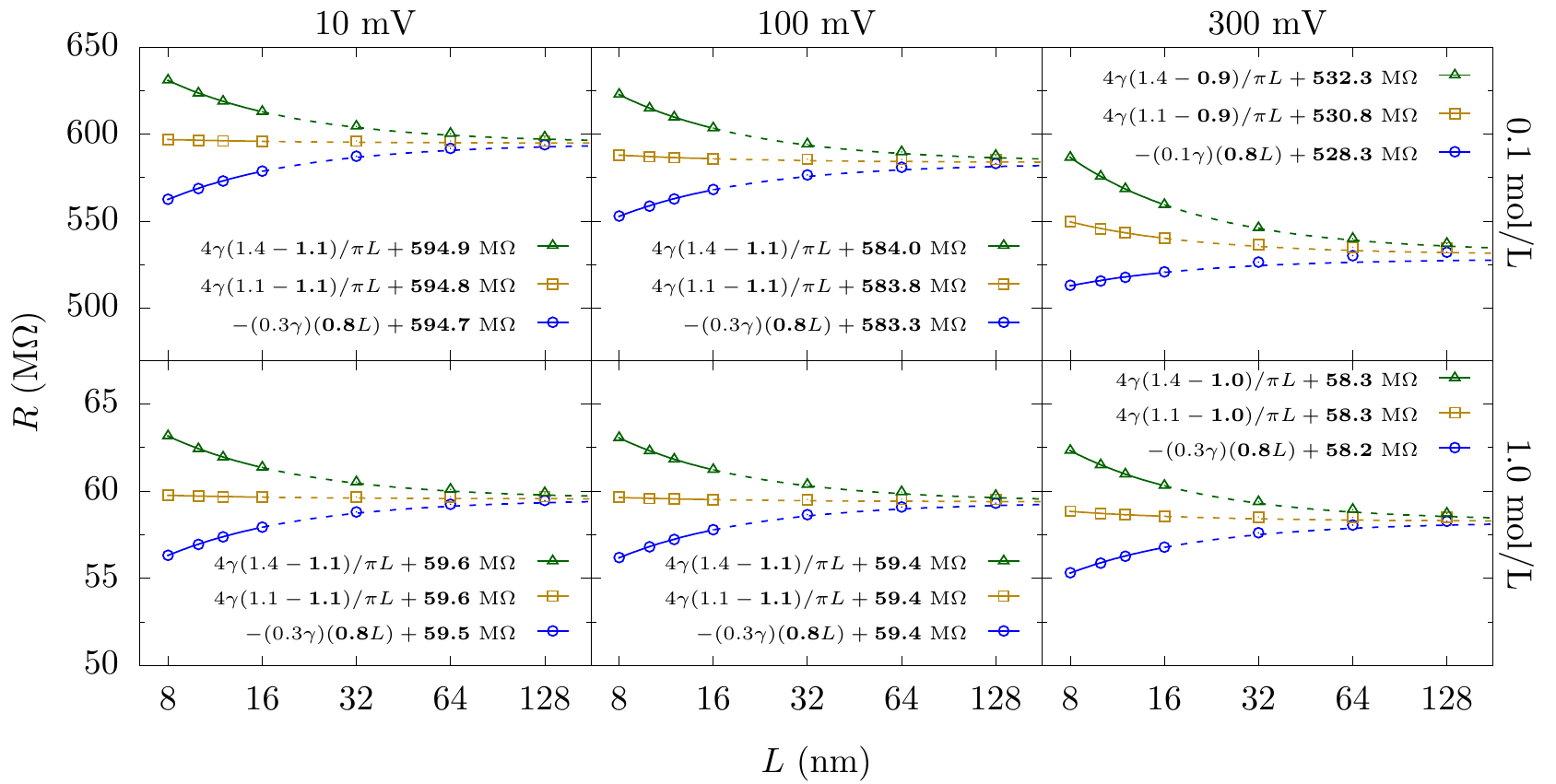} 
\caption{\label{Fig6} PNP solution of the resistance versus $L$ for cylindrical cells with $h_p=1$ nm, $a=1$ nm, and various aspect ratios ($\alpha=$ 1.4, 1.1 and 0.8 from top to bottom). The labels at the top show the applied potential and the labels at the right show the concentration of KCl. The golden aspect ratio remains the same as in the homogeneous medium calculation for the smaller voltages (10 mV and 100 mV) and for both 1 mol/L and 0.1 mol/L concentrations. For 300 mV applied voltage, the golden aspect ratio seems to take on a lower value due to the increased concentration of ions near the membrane, which is more pronounced for low concentrations due to a diminished ability to screen and generate local fields. The legends show the fits to Eq.~\eqref{golden} with the fit parameters shown in bold. The error of the fits for $f$'s and $R_\infty$'s are about 0.5 \% and 0.1 \% respectively.}
\end{figure*}

\section{RESULTS}
We first test the scaling form, Eq.~\ref{golden}, for a homogeneous medium by solving Laplace's equation for the electric potential. All simulations take a silica membrane with dielectric constant 2.1. With the rectangular cell, we compute the resistance using both finite and periodic cells (as is common in all-atom molecular dynamics). Both boundary conditions give the same results and thus we show results only for the finite cells for the continuum case (for the all-atom simulations, we have a periodic rectangular cell). Fig.~\ref{Fig3} shows that, for both cylindrical and rectangular cells, the resistance increases with $L$ for small aspect ratios and decreases for large aspect ratios as predicted from Eq.~\ref{golden}. Physically, we can understand this as follows: When the aspect ratio ($H/L$) is small, the access-like region covers most of the height of the cell. Thus, when increasing $L$, the access contribution increases, as the finite size correction -- a negative correction -- in Eq.~\ref{access} is being eliminated. On the other hand, when the aspect ratio  ($H/L$) is large, the access-like region is localized near the pore and the bulk-like region covers the remainder of the height of the cell. This bulk-like region gives a large positive contribution to the resistance, but is decreasing with $L$ (due to the fact that it decreases with the area). This transition from increasing to decreasing resistance with $L$ suggests that there should be a constant $R$ at some special value of the aspect ratio. At this special value, the two competing effects cancel. Fig.~\ref{Fig3} shows that there is an aspect ratio that shows little variation as $L$ increases. Thus, indeed, the golden aspect ratio exists and is approximately $\ga = 1.07$ for a cylindrical cell and $\ga = 1.2$ for a rectangular cell, in line with the estimate in Eq.~\ref{GASest}.

In Fig.~\ref{Fig4}(a), we examine three pore sizes and use a cylindrical cell with aspect ratio $\alpha=1.07$ where we get a visually flat profile of $R$ versus $L$. An inappropriately chosen aspect ratio, however,  will have $R$ dependent on $L$ and disagree with the classical form of the access resistance. Fig.~\ref{Fig4}(b) shows that, around $\alpha=1.07$ and at finite $L$ (32 nm), the form $R= \gamma \left( 1/2a +h_p/\pi a^2\right)+\Delta R $, fits the data with $\Delta R\approx 0 $, thus coinciding with the classical form. For $\alpha=2$, a reasonable value from a computational standpoint, the fit gives $\Delta R \approx 3$ M$\Omega$, which is an error of $\approx 30$ \% in $R_\infty$ at $a=4$ nm. 

Notice that we obtain  $h_p=1.05$ nm from the fit, which is different than the 1 nm membrane thickness. The expression $\gamma h_p/\pi a^2$ assumes the potential surfaces are flat within the full length of the pore. However, in actual pores, these surfaces are not perfectly flat, especially near the pore mouth, as is visible in Fig.~\ref{Fig1}. This difference in curvature will give a correction to $\gamma h_p/\pi a^2$ and it seems likely this is responsible for the slight difference in $h_p$. This correction becomes negligible for thick membranes, as shown in Fig.~\ref{Fig5}, where the correction comes out to be 1.0 \% for a 4 nm membrane (compared to 5 \% for the 1 nm membrane). 

To test the finite-size effect of the bulk, we can remove this source of ambiguity completely by examining the resistance on a half cylinder -- between the pore mouth set at potential $V=0$ and an electrode at one end of the cylinder. Fig.~\ref{Fig4}(c) shows this resistance versus pore radius. Since the half cylinder does not have the pore resistance, we fit it to $R= \gamma/4a +\Delta R$. Once again we get $\Delta R\approx 0$ for the aspect ratio $\alpha=1.07$ and a value substantially different than zero for other $\alpha$. These calculations, using Laplace's equation, pin the golden aspect ratio to be around $\ga = 1.07$ for the cylindrical box (and around 1.2 for a rectangular box). Furthermore, these values of the golden aspect ratio are remarkably close to the estimate in Eq.~\ref{GASest}. We also want to examine more realistic simulations, and thus we now examine the PNP equations, which are widely employed to compute ion channel behavior~\cite{eisenberg1996computing, singer2009poisson, zheng2011poisson, burger2011inverse, modi2012computational, lin2013poisson} and MD simulations.

\begin{figure}
\includegraphics{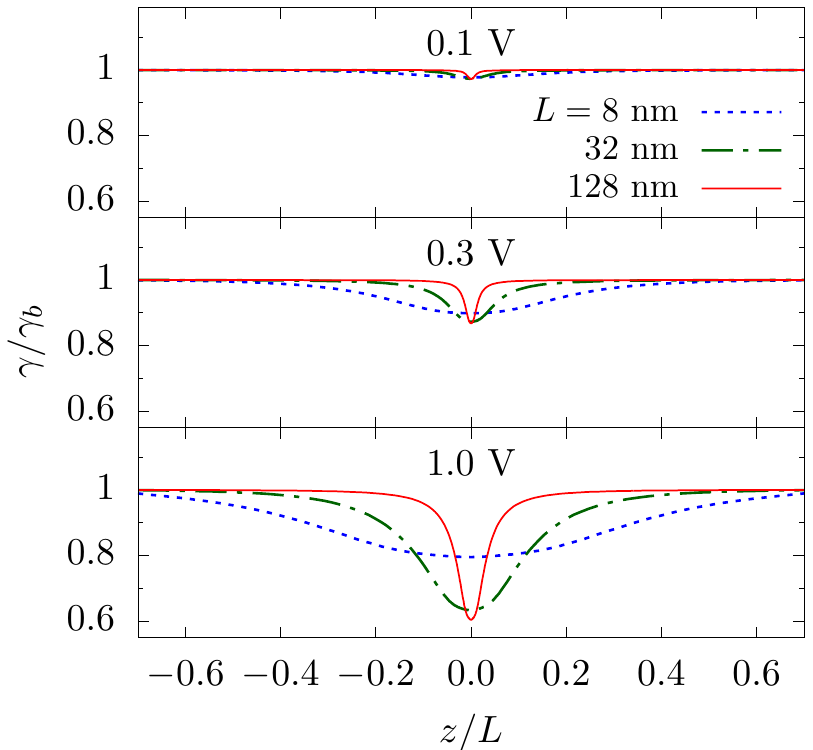} 
\caption{\label{Fig7} The resistivity along the $z$-axis for various applied potentials and cell dimensions $L$. The pore radius is $a=1$ nm and the KCl concentration is $0.1$ mol/L ($\gamma_b=710$ M$\Omega\cdot$nm). Large applied voltages enhance the density of charge carriers near the pore, decreasing the resistivity. Since the access resistance contribution is largest near the pore, see Eq.~\ref{access}, this enhancement lowers the overall access resistance and gives an apparent depression to the golden aspect ratio when examining small simulation cells.}
\end{figure}

\begin{figure}
\center
\includegraphics{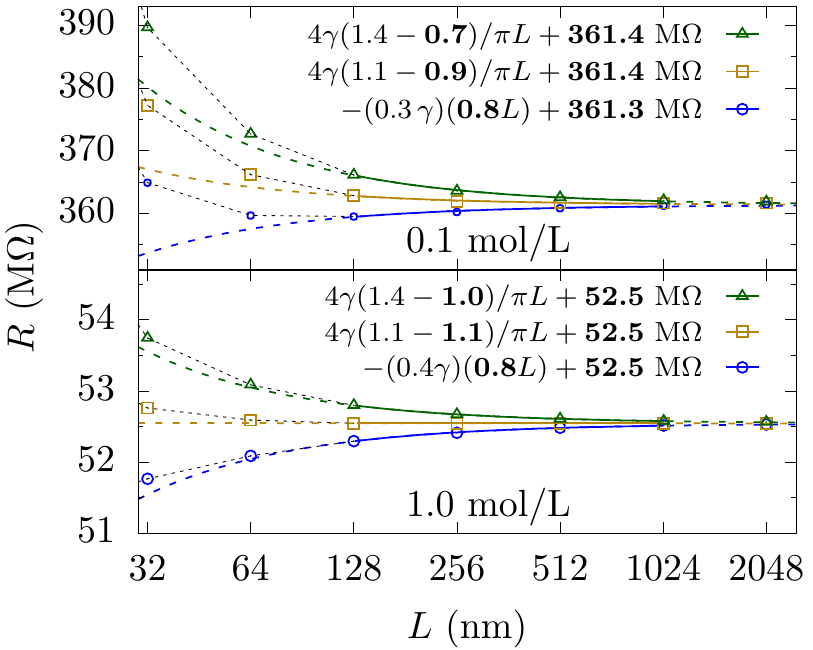} 
\caption{\label{Fig8} PNP solution of the resistance versus $L$ at applied potential of 1 V and KCl concentrations 0.1 mol/L (top panel) and 1.0 mol/L (bottom panel) for cylindrical cells with $h_p=1$ nm, $a=1$ nm, and aspect ratios ($\alpha=$1.4, 1.1 and 0.8). The solid lines show the fit with $f$ and $R_\infty$ as fitting parameters (shown in bold) and the dashed lines show the extrapolation. Since the applied field is large, the resistance at smaller $L$ ($<$ 100 nm) is highly nonlinear and cannot be scaled to extract $R_\infty$ with Eq.~\ref{golden} as is done for the smaller applied voltages in Fig.~\ref{Fig6}. Hence, a larger $L$ is necessary to employ the scaling form when the electric field is large and ion concentrations are small. Note that the resistance for the 0.1 mol/L solution is substantially smaller than at lower voltages. For 0.1 mol/L solution, the error of the fit for $f$ and $R_\infty$ are about 3 \% and 0.03 \% respectively, whereas for 1.0 mol/L solution the respective errors are less than 1 \% and 0.01 \%.}
\end{figure} 

$\,$

\noindent {\bf PNP Solution} \\
We test Eqs.~\ref{golden} for PNP simulations with two different concentrations, 0.1 mol/L and 1.0 mol/L of KCl solution, and applied voltages of 10 mV, 100 mV and 300 mV. Fig.~\ref{Fig6} shows that for 10 mV and 100 mV we obtain a similar golden aspect ratio as for the homogeneous medium solution. However, this special ratio seems to decrease for the larger voltage of 300 mV. Clearly, there are nonlinear effects coming into play when the voltage increases and/or the concentration is insufficient to screen the field without substantially perturbing the medium. We will show that this decrease is just an apparent decrease due to non-scaling [in the context of Eq.~\ref{golden}] finite-size effects. Larger $L$ will restore the scaling form. 

We first want to understand the origin of these effects, which requires that we consider the formation of concentration gradients~\cite{levadny1998}. As seen in Fig.~\ref{Fig7}, the resistivity ($\gamma =1/\sum c_\nu \mu_\nu$) is fairly constant along the axis of the simulation cell for small voltages and thus we get the same result as in the homogeneous case. For higher voltages, however, $\gamma$ is smaller near the pore as a result of classical Wien effect \cite{robinson2002,onsager1957wien}. However, the scaling form, Eq.~\ref{golden}, can still be employed as long as the resistivities of different regions do not change drastically. Relaxing the homogeneity assumption that went into the derivation of Eq.~\ref{golden}, we set $\gamma_b$, $\gamma_a$, and $\gamma_p$ to be the resistivity in the bulk-like region, the access-like region, and in the pore, respectively. Then, the total resistance (for $H \gtrsim L$) is 
\begin{align}\label{scalingPNP}
R =&R_\mathrm{access}+ R_\mathrm{bulk}+ R_\mathrm{intermediate} + \RP \nonumber\\
=& 2\left(\frac{\gamma_a}{4a} - \frac{\gamma_a}{\pi f_1 L} \right) + \gamma_b \left( \frac{\alpha L-f_2L}{\qg L^2} \right)+\frac{\gamma_b f_3}{\qg L} + \frac{\gamma_p h_p}{\pi a^2} \nonumber\\
%=& \frac{\gamma_b}{\qg L} \left( \alpha-f_1 -\frac{2\qg \gamma_a }{\pi f_1 \gamma_b}+f_3 \right)+ \left(\frac{\gamma_a}{2a}+\frac{\gamma_p h_p}{\pi a^2}\right) \nonumber\\
=& \frac{\gamma_b}{\qg L} \left[ \alpha-f\left(\frac{\gamma_a}{\gamma_b}\right) \right]+R'_\infty,
\end{align} 
where $R'_\infty=\gamma_a/2a+\gamma_p h_p/\pi a^2$ and 
\begin{equation}
f\left(\frac{\gamma_a}{\gamma_b}\right)=\frac{2\qg \gamma_a }{\pi f_1 \gamma_b}+f_2 -f_3 .
\end{equation}
Within this simplified -- ``compartmentalized'' -- inhomogeneous system, the resistance of the infinite system and the golden aspect ratio are linear function of $\gamma_a$, decreasing as $\gamma_a$ decreases. This qualitatively explains why we see the apparent decrease in the golden aspect ratio -- it is the depression of the access contribution that decreases the required height of the bulk-like region that ``balances'' it.  Of course, the resistivity is not constant throughout the access-like region and, at very high voltage and low concentration, is not even in the bulk-like region. Thus, this simplified inhomogeneous model will not accurately determine the apparent golden aspect ratio. However, it does not need to: The extensive decrease in resistivity is due to the fact that the simulation cells that are really small can not sustain large potential drops without a global perturbation to the medium. Examining much larger cells shows that the scaling form is restored, see Fig.~\ref{Fig8}, which is due to the fact that the perturbation to the medium becomes localized (in dimensionless terms) around the pore, see the curves for different $L$ in Fig.~\ref{Fig7}. 

In other words, there are finite-size effects and there are finite-size effects: The perturbation to the medium is a non-scaling finite size effect and the simulation cells need to be sufficiently large to accommodate the perturbed region. An alternative scaling ansatz could potentially be developed to handle such cases, e.g., including the perturbing effect of local fields. Its applicability would be more limited (unlike the general forms, Eq.~\ref{golden} or Eq.~\ref{constH}, which, however, may require larger $L$ to reach the scaling regime) but still helpful. Finite-size effects related to truncating the access region or asymmetric dimensions can be removed via Eq.~\ref{golden} or Eq.~\ref{constH} (for $H \gtrsim L/f$), respectively. 

$\,$

\begin{figure}
\center
\includegraphics{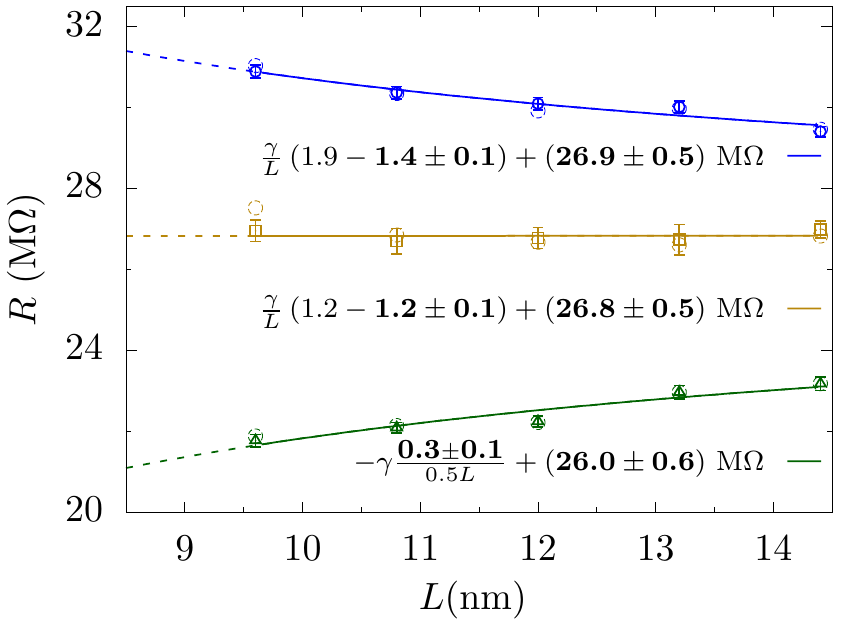} 
\caption{\label{Fig9} MD results of the resistance versus $L$ at an applied bias of 1 V and KCl concentration 1.0 mol/L, $a=1.81$ nm, and aspect ratios 1.9, 1.2 and 0.5 (from top to bottom). The raw resistance from MD is shown as open circles. Due to the equilibration, the final cell aspect ratios are not exactly 1.9, 1.2, and 0.5. The slight change in height can be corrected for in order to set $H=\alpha L+h_p$. The data with error bars show this slightly corrected data. The error bars represent $\pm$ 1 block standard error, see Ref.~\cite{grossfield2009}, which reflects error due to finite sampling time. The legends show the fits to Eq.~\eqref{golden} with the fit parameters shown in bold.  The golden aspect ratio for MD is $\approx 1.2$, which is the same as that from continuum simulations (for a rectangular cell). The scaling and/or golden aspect ratio simulations yield a resistance in agreement with the classical expression, $2 \RH + \RP$, as we demonstrated with the scaling procedure only in Sahu et. al.~\cite{sahu2018maxwell}.}
\end{figure}

\noindent {\bf MD Solution} \\
We test Eqs.~\ref{golden} for MD simulations with 1 mol/L KCl, a 1.81 nm radius graphene pore, and a 1 V applied bias. Figure~\ref{Fig9} shows results for three different aspect ratios, 1.9,1.2 and 0.5, and a cross-sectional length ranging from 9.6 nm to 14.4 nm. Both equations hold for these all-atom simulations. Moreover, we obtain the same golden aspect ratio for MD simulation as we do for the continuum calculations with a rectangular simulation cell.

\section{DISCUSSION}
Ion transport properties can be difficult to compute with all-atom molecular dynamics. In the dehydration limit, for instance, large free energy barriers exist, which entails very long (microseconds or more) simulations to accumulate enough ion crossing events to determine the current with reasonable accuracy~\cite{Sahu2017NanoLett, Sahu2017Nanoscale}. This applies even at elevated voltages [simulations are often done at 3 V or higher~\cite{sathe2011}], which risks introducing nonlinearities and prohibits a direct comparison with experiment. Contextual properties of ion transport -- fluctuations of the pore or membrane, localized dipoles or charges, van der Waals interactions and dehydration, geometrically ``imperfect'' pore shapes, etc. -- make these simulations even more difficult to understand and control (e.g., slow fluctuations of pore structure require long times to acquire the requisite statistics). To capture these effects, therefore, one would like to keep the simulation size as small as possible but without introducing excessive errors into the computation. 

We have developed a scaling theory to extract both the pore and access contributions to resistance without going to excessively large simulation cells (i.e., $L<16$ nm for linear response). This theory was originally employed to determine the access contribution -- including how to properly define the pore size in the presence of contextual properties -- to graphene nanopores with all-atom MD~\cite{sahu2018maxwell}. While this does not correct for force-field effects and related issues (sampling, conformational changes, etc.), it does solve for the geometry/dimensions of the setup. Using this theory, we demonstrated that the ``golden aspect ratio'' -- a special aspect ratio where finite size effects are eliminated -- exists in both continuum and all-atom simulations (and we expect in Brownian dynamics as well). Previously, we only used an aspect ratio close to an initial estimate of the golden aspect ratio (and, indeed, we had little variation of $R$ with $L$) but large enough to ensure we decrease from above to $R_\infty$~\cite{sahu2018maxwell}. 

The golden aspect ratio gives a completely flat (i.e., without non-monotonic or oscillatory behavior) resistance versus simulation cell size. The value for the golden aspect ratio is about $\ga = 1.07$ for cylindrical cells and 1.2 for rectangular (for both continuum and all-atom simulations). It is relatively insensitive to voltages and ion concentrations within experimentally relevant regimes (i.e., around 0.1 V applied voltages and 0.1 mol/L to 1.0 mol/L concentrations). For large voltages and small concentrations, simulation cells must be sufficiently large to remove non-scaling finite-size effects, after which the scaling and the golden aspect ratio are restored. 

This scaling approach will be most useful when examining pores that do not have very high pore resistances -- due to dehydration or a long length channel -- so that the access resistance becomes a dominant or equal contributor rather than a correction to the resistance. Graphene and other atomically thin pores have a dominant or substantial access contribution all the way down to the dehydration limit~\cite{sahu2018maxwell}, and thus the scaling approach is essential. Biological channels can fall into either category, sometimes requiring a correct determination of the access contribution. For instance, $\alpha$-hemolysin pores have a 1 G$\Omega$ resistance (at 1 mol/L KCl)~\cite{aksimentiev2005}, compared to an estimated [from Eq.~\ref{MaxwellHall}] 40 M$\Omega$ access contribution. On the other hand, a sodium channel of radius 0.3 nm and effective length 0.5 nm has both an access and pore resistance that are nearly equal~\cite{hille1968}. In low salt concentrations, access resistance may become the dominating resistance in a wide variety of cases~\cite{alcaraz2017ion}. While we have not simulated a biological ion channel here, a direct application of the golden aspect ratio would take the simulation cell height to be $1.2\cdot L + h_p$, with $h_p$ the channel length. For instance, for $\alpha$-hemolysin as a model biological channel, $h_p\approx 10$ nm, the simulation cell height should be roughly on the order of 27 nm (for the cross-sectional dimension to be reasonably larger than channel width and protein assembly, $L$ should be about 14 nm~\cite{aksimentiev2005}). However, this idea requires testing with different biological channels, as in many cases (including $\alpha$-hemolysin), the membrane thickness and channel length are not the same. This distorts the equipotential surfaces. Given the ease with which the scaling approach and golden aspect ratio can be employed (e.g., setting $H \approx 1.2\cdot L + h_p$ and simulating a handful of different $L$), it is reasonable to use it first even in the cases the where access resistance is a correction.

Care must be taken, though, to work with sufficiently large $L$ in simulations, especially all-atom MD. For instance, a simulation cell that is too small may not have enough ions to properly screen localized dipoles or charges (or otherwise give a cell boundary too close to these local electrostatic variations, preventing proper Debye screening). Similarly, the partitioning of the voltage drop across the cell can yield a strong field across the pore for small simulation cells (as there is less voltage drop in the bulk medium), which can introduce non-linearities in the transport properties. This further supports the use of the scaling analysis, as it will also allow one to identify these ``non-scaling'' finite size effects and eliminate them when going to large enough simulation cells. Moreover, it goes without saying that computations should be done as close to experimental and biologically-relevant conditions as possible. Our work gives new reasons to do simulations in the linear response limit, as it is only then that some non-scaling finite size effects are negligible for simulation cell sizes in the 16 nm regime.

In addition, the classical form of access resistance assumes an infinite bulk, which fails for micro- and nano-scale systems. In particular, if one has a narrow (micro- or nano-fluidic) constriction leading up to a membrane/pore or some other ``active'' region, then access resistance no longer follows Hall's expression, but rather has to be corrected for the geometric setup \cite{green2016}. Our approach gives a method to map simulations to experimental geometries (or vice versa), thus allowing for contextual properties to be simulated with reasonable computational power. 

The primary power of MD is to study contextual properties of pores, such as the effect of atomic/molecular fluctuations of the pore/membrane and the presence of (partial) charges, but requires extensive computational resources. The golden aspect ratio and scaling approaches will allow the quantitative extraction of both pore and access contributions to the resistance with minimal resources. Its use has already shed light on how to define pore size for geometrically imperfect pores, including edge fluctuations and dehydration/van der Waals interactions~\cite{sahu2018maxwell}. By extension, this will give the first calculated values of access resistance in the presence of, e.g., molecular-scale/protein fluctuations in biological ion channels and other contextual properties (charges/dipoles) and open a new era in comparing computed and measured values of resistance. \\

%\section*{Appendix A: Access resistance in a finite region}
\appendix*{\bf Appendix: Access resistance in a finite region}\\

Here, we calculate access resistance for a finite size region while keeping the ellipsoidal symmetry. This problem can be solved using the  rotational elliptic coordinates. We begin with oblate spheroidal coordinates, ($\mu$, $\nu$, $\phi$), which are defined in terms of the Cartesian coordiates as
\begin{align}
x &= a \cosh\mu\, \cos\nu\, \cos\phi \nonumber \\
y &= a \cosh\mu\, \cos\nu\, \sin\phi \nonumber \\
z &= a \sinh\mu\, \sin\nu\,
\end{align}
 Rotational elliptic coordinates, $\xi=\sinh\mu $ and $\eta=\sin\nu$, are related to the cylindrical coordinates as
\begin{align}
z= \rp \xi \eta \nonumber \\
\rho = \rp \sqrt{(1+\xi^2)(1-\eta^2)} ,
\end{align}
where $\rp$ is the radius of the pore. Laplace's equation in this coordinate system is 
\begin{equation}
\frac{\partial }{\partial \xi} \left[ (1+\xi^2) \frac{\partial V}{\partial \xi} \right]+\frac{\partial }{\partial \eta} \left[ (1-\eta^2) \frac{\partial V}{\partial \eta} \right]=0
\end{equation}
This equation can be solved via separation of variables, $V=H(\eta) \Xi (\xi)$, which gives 
\begin{equation} \label{xi}
\frac{\partial }{\partial \xi} \left[ (1+\xi^2) \frac{\partial \Xi}{\partial \xi} \right] + \lambda \,\Xi = 0 
\end{equation}
and
\begin{equation} \label{eta}
\frac{\partial }{\partial \eta} \left[ (1-\eta^2) \frac{\partial H}{\partial \eta} \right]-\lambda\, H =0 ,
\end{equation}
where $\lambda$ is a constant. The boundary condition (ii), $\partial V /\partial \eta=0$ at $\eta=0$, yields $\lambda=0$. Integrating Eq.~\ref{xi} thus gives 
\begin{equation}
V= H\, \Xi = H\, \int \frac{b_1}{1+\xi^2} d \xi =H\, b_1 \tan^{-1}\xi+b_2 ,
\end{equation}
where $b_1$ and $b_2$ are the constants of integration.  Using the boundary condition (i), $V(\xi=0) = 0$, we get $b_2=0$.
This simplifies the potential to $V=H\, b_1 \tan^{-1}\xi$.
The boundary condition (iii), $V(\xi=l/\rp) = V_0 $, gives $V_0 = H\, b_1 \tan^{-1}(l/\rp)$.
Taking the ratio of these two expressions gives Eq.~\ref{Potential} of the main text, 
\begin{equation}
\frac{V}{V_0} =  \frac{\tan^{-1}\xi}{\tan^{-1}(l/\rp)} . \label{PotApp}
\end{equation}
The current through the pore is 
\begin{align}
I =& \frac{2\pi}{\gamma} \int_0^{\rp} \frac{\partial V}{\partial z}\Big|_{z=0}  \rho d\rho =  \frac{2 \pi}{\gamma}\int_0^{\rp} \frac{1}{\rp \eta} \frac{\partial V}{\partial \xi}\Big|_{\xi=0}\ \rho d\rho  \nonumber \\
%=&  \frac{2 \pi V_0}{\gamma \tan^{-1} (l/\rp)} \int_0^{\rp} \frac{1}{\sqrt{\rp^2-\rho^2}}  \rho d\rho  \nonumber \\
=& \frac{2 \pi  \rp V_o}{\gamma \tan^{-1} (l/\rp)} . \label{Current}
\end{align}
Equations~\ref{PotApp} and~\ref{Current} give Eq.~\ref{access}. 
\begin{align}
\RA &= \frac{\gamma\tan^{-1}(l/\rp)}{  2 \pi \rp}\nonumber \approx  \RH -\frac{\gamma}{2\pi l}.\nonumber
\end{align}

\section*{Acknowledgments}
We thank J. Elenewski, D. Gruss, and C. Rohmann for helpful comments. S. S. acknowledges support under the Cooperative Research Agreement between the University of Maryland and the National Institute of Standards and Technology Center for Nanoscale Science and Technology, Award  70NANB14H209, through the University of Maryland.

%\bibliography{reference} 


\end{document}